\documentclass[manuscript]{aastex}

\usepackage{color}
\usepackage{amsmath}
\usepackage{subfigure}

\shorttitle{Stress and Failure Analysis of (29075) 1950 DA}
\shortauthors{Hirabayashi and Scheeres}

\begin{document}


\title{Stress and Failure Analysis of Rapidly Rotating Asteroid (29075) 1950 DA}


\author{Masatoshi Hirabayashi\altaffilmark{1}}
\email{masatoshi.hirabayashi@colorado.edu}

\author{Daniel J. Scheeres\altaffilmark{1}}
\affil{Aerospace Engineering Sciences, 429 UCB, University of Colorado, Boulder, CO 80309-0429 United States}


\altaffiltext{1}{Research Associate, Colorado Center for Astrodynamics Research, 
Aerospace Engineering Sciences, University of Colorado Boulder}

\altaffiltext{2}{A. Richard Seebass Chair, Professor, Colorado Center for Astrodynamics Research, 
Aerospace Engineering Sciences, University of Colorado Boulder}


\begin{abstract}
\cite{Rozitis2014} recently reported that near-Earth asteroid (29075) 1950 DA, whose bulk density ranges from 1.0 g/cm$^3$ to 2.4 g/cm$^3$, is a rubble pile and requires a cohesive strength of at least 44 Pa to 74 Pa to keep from failing due to its  fast spin period. Since their technique for giving failure conditions required the averaged stress over the whole volume, it discarded information about the asteroid's failure mode and internal stress condition. This paper develops a finite element model and revisits the stress and failure analysis of 1950 DA. For the modeling, we do not consider material-hardening and softening. Under the assumption of an associated flow rule and uniform material distribution, we identify the deformation process of 1950 DA when its constant cohesion reaches the lowest value that keeps its current shape. The results show that to avoid structural failure the internal core requires a cohesive strength of at least 75 Pa - 85 Pa. It suggests that for the failure mode of this body, the internal core first fails structurally, followed by the surface region. This implies that if cohesion is constant over the whole volume, the equatorial ridge of 1950 DA results from a material flow going outward along the equatorial plane in the internal core, but not from a landslide as has been hypothesized. This has additional implications for the likely density of the interior of the body. 
\end{abstract}


\keywords{minor planets, asteroids}



\section{Introduction}
Near-Earth asteroid (29075) 1950 DA is currently considered to be among the most hazardous asteroids due to its close encounter to the Earth in 2880 \citep{Giorgini2002}. \cite{Busch2007} conducted comprehensive radar observations of this asteroid and derived two possible shape models. The first is a prograde model, which is a spherical shape, and the second is a retrograde model, which is an oblate shape. \cite{Farnocchia2014} reported that the orbital semi-major axis of 1950 DA has been changing due to the Yarkovsky effect and confirmed that the retrograde model was consistent with their analysis. Because of its spin period, 2.1216 hr, this object may be close to its structural failure point if this body is a rubble pile. If 1950 DA has no cohesion, i.e., zero shear resistance at a zero-normal stress, the bulk density should be higher than 3.5 g/cm$^3$ to prevent the body from failing structurally \citep{Busch2007}. Using an advanced thermophysical model and archival WISE thermal-infrared data to derive a bulk density of $1.7 \pm 0.7$ g/cm$^3$, \cite{Rozitis2014} indicated that 1950 DA should have cohesive strength to keep its current shape. They used Holsapple's limit analysis technique to derive the lowest cohesion for keeping the current shape of 1950 DA. The detailed development was described by \cite{Holsapple2004,Holsapple2007}. However, since the applied technique was based on taking the average of stress components over the whole volume, this discards crucial information about the disruption processes. 

Disruption events of asteroids have been recently reported. For example, active asteroid P/2013 R3 was observed breaking into multiple components \citep{Jewitt2014}. From this fact, \cite{Hirabayashi2014_R3} found that this asteroid has a level of cohesion ranging between 40 Pa and 210 Pa. Other disrupting bodies have also been observed. Active astroid P/2013 P5 is currently shedding discontinuous dust tails \citep{Jewitt2013}. \cite{Sheppard2014} discovered a new active asteroid, (62412) 2000 SY178, exhibiting a tail from its nucleus. These observations reveal that disruption events in our solar system are more common than previously known. Thus, a understanding of such disruption events provides better constraints on the formation of small bodies. 

\cite{Holsapple2008A} developed a finite element model that takes into account plastic deformation characterized by the von Mises yield criterion, a pure-shear-dependent criterion and applied it to the failure condition of a rotating, non-gravitating ellipsoid. We extend his model to a model that can take into account self-gravity and plastic deformation characterized by a shear-pressure yield criterion. For our computations we use a commercial finite element software, ANSYS, version 15.03. We investigate the internal condition of the 1950 DA retrograde model at the current spin period. Our analysis provides a more precise lower bound on the necessary cohesion to hold 1950 DA together in a stable state. More importantly, our stress analysis also shows that the preferred mode of failure is for the central region of the body to collapse along the rotation axis, which would cause the equator to expand outward. The analysis indicates that the body's failure state is not close to either surface material being shed or landsliding. Given that plastic flow will generally increase volume in an associated flow rule \citep{Chen1988}, this failure mode predicts that the central core of the asteroid may have a reduced density, which is a previously unpredicted state for such oblate, rapid rotators. 

\section{Physical Properties of 1950 DA}
\label{sec:phyProp}
Information from radar observations can constrain an asteroid's near-surface bulk density, as a function of the material density and porosity \citep{Magri2001}. Using this technique, \cite{Busch2007} obtained the minimum surface material density as 2.4 g/cm$^3$ and derived two compositional possibilities of the surface, enstatite chondrite and nickel-iron. On the other hand, the thermophysical model used by \cite{Rozitis2014} gave a bulk density of $1.7 \pm 0.7$ g/cm$^3$. Here, we use the bulk density, derived by \cite{Rozitis2014}, which also includes the estimated value by \cite{Busch2007}. 

The retrograde model of 1950 DA derived by \cite{Busch2007} is oblate and has an equatorial ridge (Fig. 5 in their article). In the following discussion, the minimum, intermediate, and maximum moment of inertia axes are defined as the $x_1$, $x_2$, and $x_3$ axes, respectively. The mesh is constructed using the \cite{Busch2007} retrograde model. Each element is a 10-node tetrahedron. The numbers of elements and nodes are 5569 and 8595, respectively. The initial size of the shape is fixed to be constant and so the initial volume is 1.145 km$^3$. The spin period is 2.1216 hr. It is assumed that 1950 DA is a principal axis rotator and is affected only by self-gravity and rotation. Here, the rotation pole is fixed along the $x_3$ axis. We define Young's modulus as $1.0 \times 10^7$ Pa, a typical value for geological sand and gravel, and Poisson's ratio as 0.25, giving compressibility of the volume. A friction angle, $\phi$, is fixed at 35 degrees, the mean value of a friction angle of a geological material ranging from 30 degrees and 40 degrees \citep{Lambe1969}. Cohesion is kept as a free parameter and we will determine the lowest cohesion that prevents this object from failing structurally. Here, we calculate finite element solutions for three different bulk density cases: 1.0 g/cm$^3$, 1.7 g/cm$^3$ and 2.4 g/cm$^3$. 

\section{Finite Element Modeling}
\label{Sec:FEM}
In the current finite element model, if a stress state is below the yield condition, the behavior of a material follows linear elasticity. It is assumed that the behavior of plastic deformation does not include material-hardening and softening (perfect plasticity). Body forces are defined at each node so that the total force acting on each element is equally split and apparently concentrates on its nodes. In nature, an asteroid rotates in free space, implying that there are no boundary conditions. However, in our numerical environment, the setting of body forces makes it difficult for the simulations to converge correctly because it could cause rigid motion, i.e., rotation and translation. To avoid this issue we artificially constrain six degrees of freedom to eliminate such motion. 

1950 DA is currently considered to be a rubble pile \citep{Rozitis2014}. For such a body, the mechanical behavior of its material depends on shear and pressure \citep{Lambe1969,Jaeger1976}. In this study, we use the Drucker-Prager yield criterion, a smooth function in the principal stress space, which is given as 
\begin{eqnarray}
f (I_1, J_2) = \alpha I_1 + \sqrt{J_2} - k = 0,  
\end{eqnarray}
where 
\begin{eqnarray}
I_1 &=& \sigma_1 + \sigma_2 + \sigma_3, \\
J_2 &=& \frac{1}{6} [(\sigma_1 - \sigma_2)^2 + (\sigma_2 - \sigma_3)^2 + (\sigma_3 - \sigma_1)^2].
\end{eqnarray}
$\sigma_1$, $\sigma_2$, and $\sigma_3$ are the principal components of the stress state, and $\alpha$ and $k$ are free parameters that need to be fixed. We choose these parameters by the following consideration. Since the retrograde shape model of 1950 DA is oblate, only the stress components along the $x_1$ and $x_2$ axes are affected by the centrifugal force. In such a case, the relation of the stress components is described as $\sigma_1 \sim \sigma_2 > \sigma_3$ (here, a negative value means compression, while a positive value indicates tension). Thus, since the actual stress state is near the compression meridian of the Mohr-Coulomb yield criterion, $\sigma_1 = \sigma_2 \ge \sigma_3$, we choose $\alpha$ and $k$ such that the Drucker-Prager yield envelope touches the Mohr-Coulomb yield envelope at this meridian. These parameters are then given as \citep{Chen1988}
\begin{eqnarray}
\alpha &=& \frac{2 \sin \phi}{\sqrt{3} (3 - \sin \phi)}, \\
k &=& \frac{6 c \cos \phi}{\sqrt(3) (3 - \sin \phi)}, \label{Eq:k}
\end{eqnarray}
where $c$ is cohesion and $\phi$ is a friction angle. 

Plastic deformation is assumed to be small. The constitutive law of plastic deformation is constructed based on an associated flow rule. This rule guarantees that the direction of plastic deformation is always perpendicular to the yield envelope. The plastic strain rate $\dot \epsilon^p_{ij}$, where $i$ and $j$ are indices, is described by the rate of an arbitrary coefficient $\lambda$ times the partial derivative of the function $f$ with respect to the stress tensor $\sigma_{ij}$, which is given as
\begin{eqnarray}
\dot \epsilon^p_{ij} = \dot \lambda \frac{\partial f(\sigma_{ij})}{\partial \sigma_{ij}}, \label{Eq:associated}
\end{eqnarray}
Note that it is known that the behavior of a typical geological material would rather follow a non-associated flow rule than be characterized by an associated flow rule. However, since (1) choosing an associated flow improves the convergence of a finite element solution in the current model and (2) consideration of a non-associated flow rule requires additional free parameters (usually, unknown), we use the associated flow rule described in Eq. (\ref{Eq:associated}). We leave the application of a non-associated flow rule as future work.

A plastic solution depends on loading path. Although such a path represents evolutional history, in general it is usually unknown. Here, we assume that the gravitational and centrifugal forces acting on small elements ramp up linearly. This load step implies that the evolution of 1950 DA results from its accretion process due to a catastrophic disruption. The time scale of the accretion process due to a catastrophic disruption is considered to be negligibly short, compared to the life time of an asteroid, probably much less than a year \citep{Michel2001, Michel2002, Michel2004}. 

To derive the plastic deformation mode at the lowest cohesion that can keep the current shape of 1950 DA, we conduct an iteration method described as follows. First, we choose a value of cohesion and calculate a plastic solution that corresponds to it. If this solution only includes elastic states everywhere, we choose a lower value for cohesion and recalculate the stress solution. We iterate this process until we obtain a solution in which a majority of the internal structure reaches the yield condition. To go beyond such a condition, one usually encounters computational issues and so we terminate our iteration. Thus, the solution described in this study is as close as possible to the numerical computation limit of plastic analysis. 

\section{Result}
We implement a 3-dimensional finite element model of 1950 DA (the retrograde model) on ANSYS, version 15.03. The purpose of this paper includes visualization of plastic deformation in the interior of an asteroid. To do so, we use the ratio of the current stress state to the yield stress, so-called ``stress ratio" \citep{ANSYSThr}. Although this parameter does not show the plastic state during a loading path, (in order words, it does not show the history of the plastic state), it is useful to visualize the final state. Using the derivation given by \cite{ANSYSThr}, we write the current stress state, the so-called ``equivalent stress", as 
\begin{eqnarray}
\sigma_e = \alpha I_1 + \sqrt{J_2}, 
\end{eqnarray}
where $I_1$ and $J_2$ are obtained using the current stress state. The yield stress is given as
\begin{eqnarray}
\sigma_y = k, 
\end{eqnarray}
where $k$ is given in Eq. (\ref{Eq:k}). Then, the stress ratio is obtained as
\begin{eqnarray}
N = \frac{\sigma_e}{\sigma_y}. \label{Eq:SRAT}
\end{eqnarray}
If $N = 1$, an element in the body is in a plastic state. 

Figure \ref{Fig:failureMode} shows a plastic solution for the case of a bulk density of 1.0 g/cm$^3$. Figure \ref{fig:FM0} gives the total deformation vectors of the shape in meters. Figure \ref{fig:FMA} shows the stress ratio over the cross section along the $x_1$ and $x_3$ axes. The cohesion for this case is 75 Pa. The stress ratio around the center region is over 0.99, while the region with a stress ratio of 0.9 is much wider than that of 0.99. This indeed indicates that plastic deformation occurs in the center of the interior first. This plastic deformation comes from the fact that the strong centrifugal force acts in the horizontal plane, while only the gravitational force acts along the $x_3$ axis. Thus, the internal core experiences strong shear, equivalent to half of a difference between the principal components of the stress state, leading to a large scale of plastic flow. Specifically, the stress components in the $x_1$ and $x_2$ axes may be strongly tensile due to the fast spin period, while the component in the $x_3$ axis is always compressive. For this reason, plastic deformation occurs in the internal core and so the shape becomes more oblate than the current shape. On the other hand, Fig. \ref{fig:FMB} indicates the stress ratio on the surface. The stress ratio of almost all the surface area is below 0.8, meaning that the near-surface region is below the yield condition. This implies that for the case of uniform material distribution, the central core of 1950 DA is the most sensitive part to structural failure, while the surface is not. Based on an associated flow rule, plastic flow must always be accompanied by an increase in volume \citep{Chen1988}. Hence, since the mass is conserved over the granular flow, the porosity of the internal core should be lower than that of other regions.

We calculate finite element solutions for three difference bulk density cases: 1.0 g/cm$^3$, 1.7 g/cm$^3$, and 2.4 g/cm$^3$. Figure \ref{Fig:densSol} gives the variation of the lowest cohesion with respect to different bulk densities by the solid line and the dark markers. The resolution of the lowest cohesion is 2 Pa. The shadow area shows the cohesion area in which the original shape can remain, while the white region indicates the prohibited cohesion for the existence of 1950 DA. The lowest cohesion ranges from 75 Pa to 85 Pa. It is also found that the lowest cohesion increases as the bulk density becomes higher. This comes from the fact that as the bulk density increases, higher body forces induce higher shear and normal stresses, requiring higher cohesion for keeping the current shape. 

\section{Discussion}
The sensitivity of the internal core of an oblate body to failure is a new insight about the internal structure of an asteroid. Earlier studies using a volume averaging method (e.g., \cite{Holsapple2007}, \cite{Hirabayashi2014}) could only derive an upper bound condition for structural failure of an asteroid, while the present technique can consider a detailed deformation mode as well as the precise condition for it. For the case of 1950 DA, since the spin state is considered to be near its failure condition, we find that at the lowest cohesion for keeping its current shape, the internal core fails structurally, while the surface region does not. Because of this, horizontal outward deformation occurs in that plastic region and pushes the outer region to build the equatorial ridge. This result implies that if cohesion of the material of 1950 DA were constant over the whole volume, the formation of the equatorial ridges would result from plastic deformation of the internal core and not from a landslide. The present failure model for 1950 DA predicts that the core of the asteroid should be ``under-dense". This is the opposite of the prediction if the surface regolith flows to the equator to form the bulge -- then the bulge is under-dense. This is a distinction that can be specifically tested for future asteroid exploration missions such as the OSIRIS-REx mission, which should be able to distinguish whether the core is under- or over-dense relative to the rest of asteroid (101955) Bennu \citep{Nolan2013, Takahashi2014}. 

\begin{figure}[ht!]
	\begin{center}
		\subfigure[]{
         		\label{fig:FM0}	
		\includegraphics[width=5in]{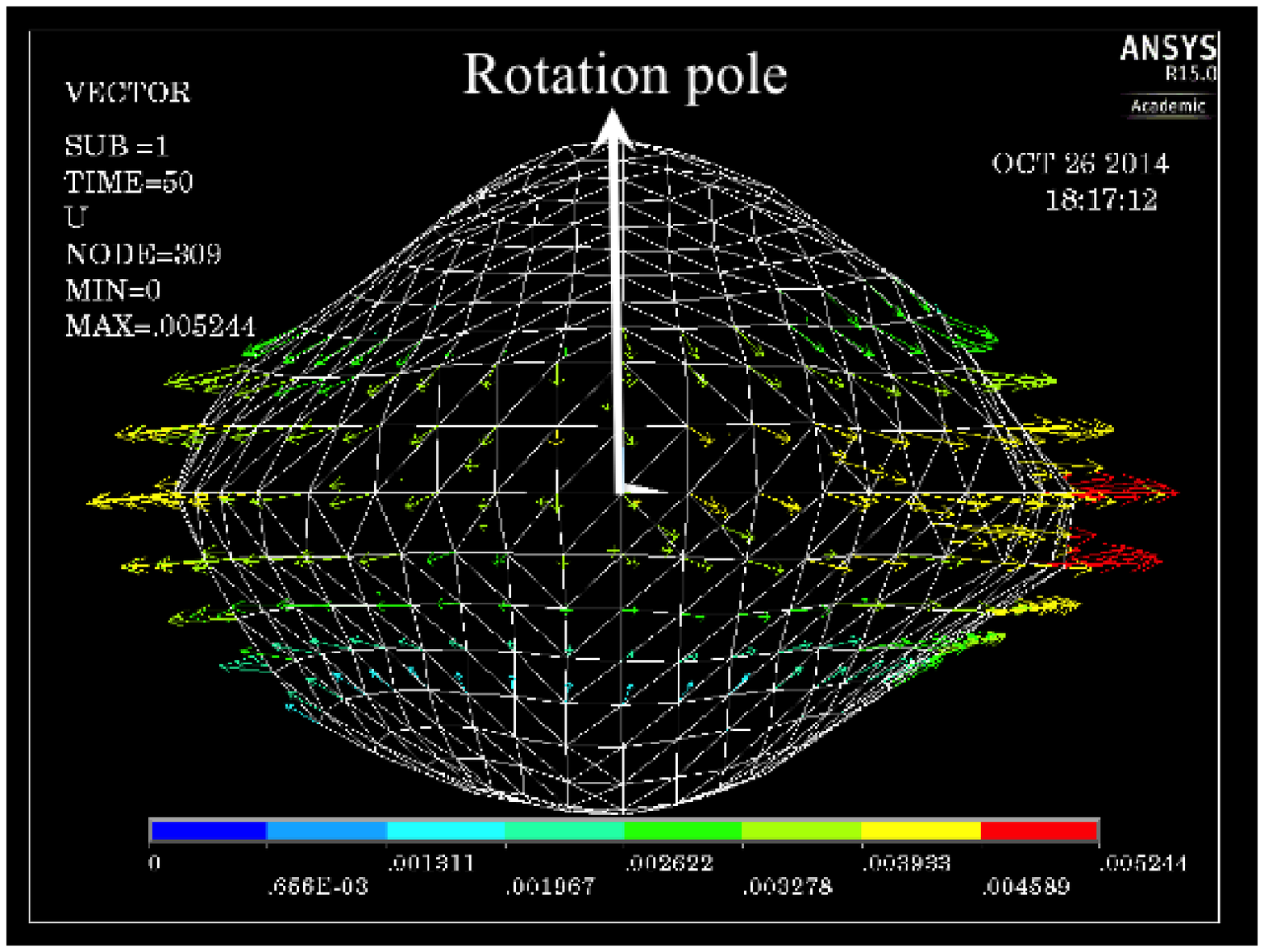}
          	} \\
		\subfigure[]{
         		\label{fig:FMA}	
		\includegraphics[width=3in]{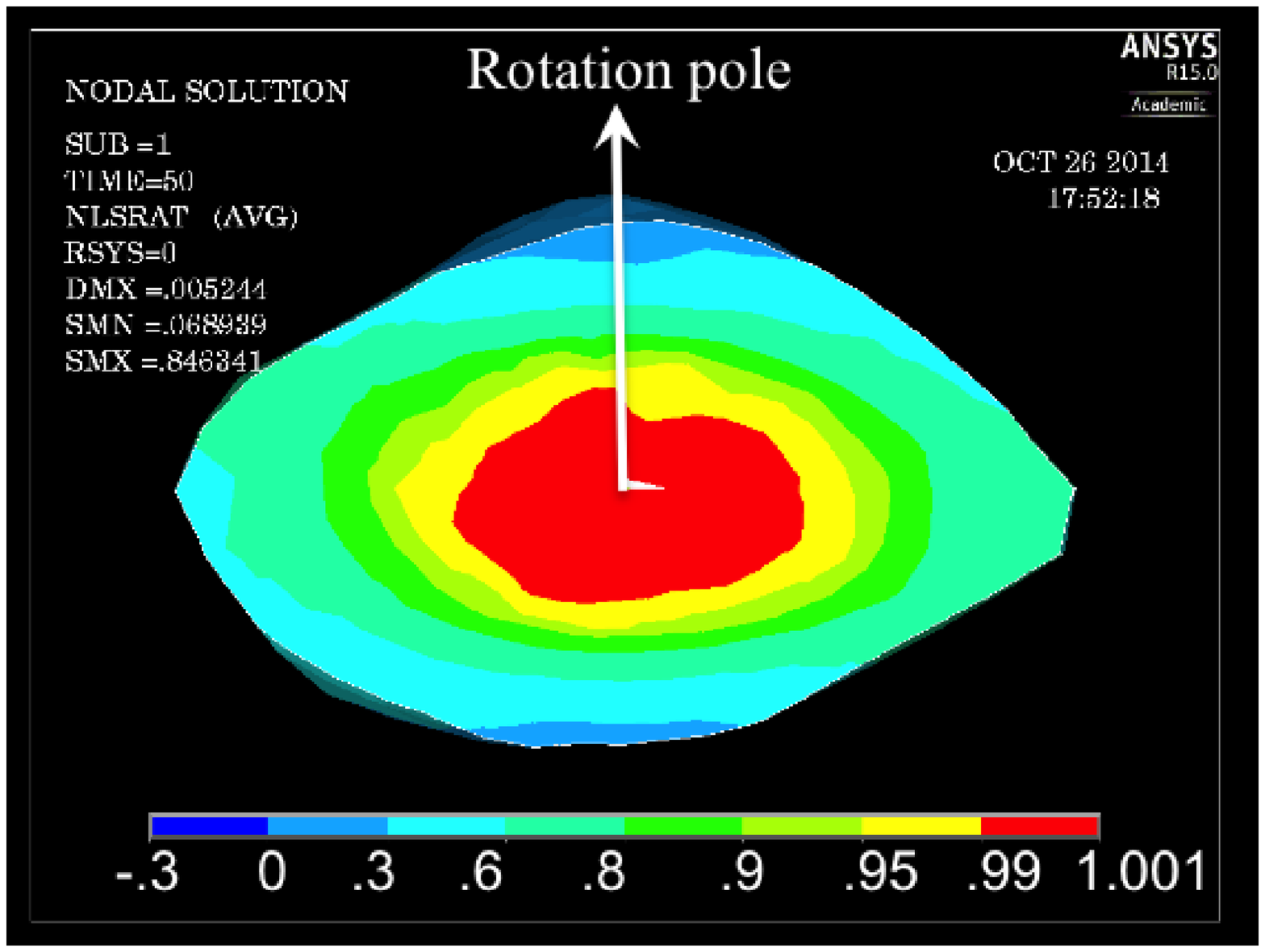}
          	}
		\subfigure[]{
         		\label{fig:FMB}	
		\includegraphics[width=3in]{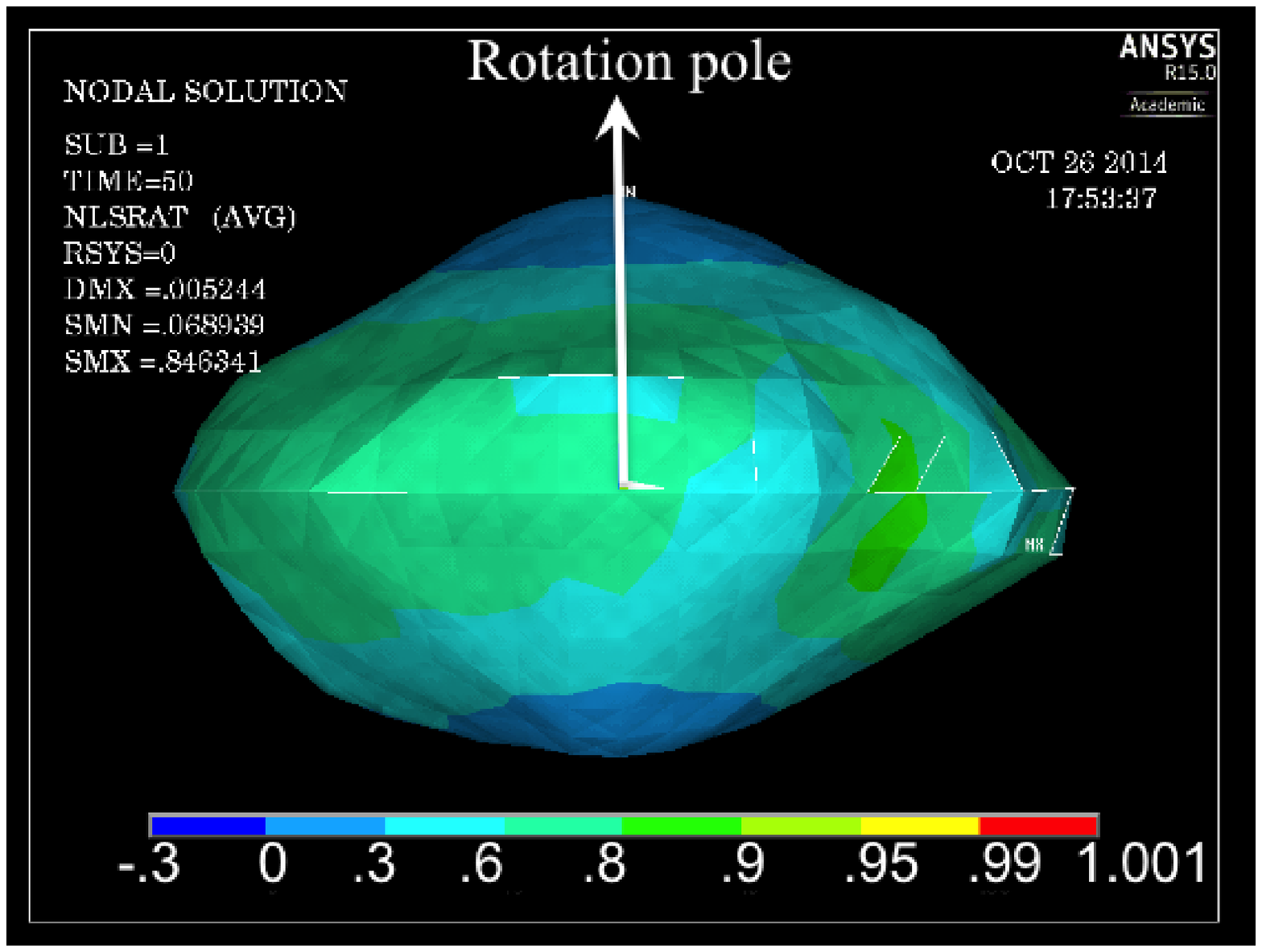}
          	} 	
	\caption{Plastic solution of 1950 DA with a bulk density of 1.0 g/cm$^3$. Figure \ref{fig:FM0} indicates the deformation vectors of the shape in meters, Fig. \ref{fig:FMA} shows the stress ratio over the cross section along the minimum and maximum moment of inertia axes, and Fig. \ref{fig:FMB} describes that on the surface. The regions whose stress ratio ranges from 0.99 to 1.001 experience plastic failure.}
	\label{Fig:failureMode}
	\end{center}
\end{figure}

\begin{figure}[hb]
  \centering
  \includegraphics[width=6in]{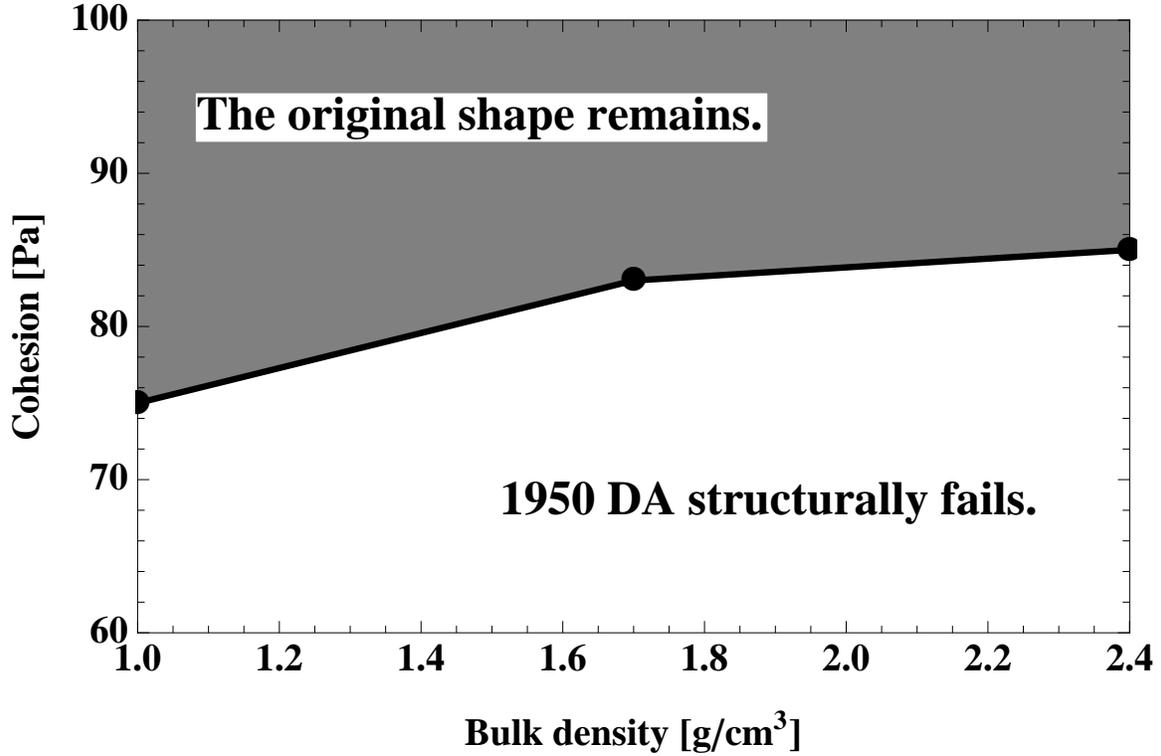}
  \caption{Lowest cohesion that prevents 1950 DA from failing structurally for three different bulk density cases: 1.0 g/cm$^3$, 1.7 g/cm$^3$, and 2.4 g/cm$^3$. The shadow region describes that the original shape can remain, while the white region indicates that it must fail structurally.}
  \label{Fig:densSol}
\end{figure}

From the current finite element model, it is unlikely that surface failure occurs. As seen in Fig. \ref{fig:FMB}, the surface region is always below the yield condition. This implies that although the centrifugal force exceeds the gravitational force, the surface region does not fail structurally and so the surface particles will not fly off because of its cohesion. Here, we examine whether or not a small boulder is stable in the case when the cohesion obtained above represents the material strength. This condition can be obtained by considering the force balance on a boulder at the equatorial region. Here, to simplify  this analysis, we assume that 1950 DA and the boulder are spherical bodies. The masses of 1950 DA and the boulder are denoted as $M$ and $m$, respectively. The radii of 1950 DA and the boulder are defined as $R$ and $r$, respectively. The bulk density of 1950 DA and the material density of the boulder are given as $\rho_B$ and $\rho_G$, respectively. Also, $\omega$ is the current spin rate of 1950 DA. In order for the boulder to separate, the centrifugal force acting on the boulder should exceed the gravitational force, which is given as
\begin{eqnarray}
m (R + r) \omega^2 \ge \frac{G M m}{(R+r)^2} + f_{cohesion}. 
\end{eqnarray}
The second term on the right-hand side indicates a cohesion force. Here, using the results in \cite{Sanchez2014}, we simply describe this term as $\sim 2 c \pi r^2$, where $c$ represents cohesion and $2 \pi r^2$  is an apparent area of the boulder touching the surface. We assume that the radius of the boulder is very small compared to that of 1950 DA, i.e., $r \ll R$. By choosing $\rho_B \sim 2.4$ g/cm$^3$, an upper bound for the bulk density of 1950 DA, $\rho_G \sim 6$ g/cm$^3$, from the fact that 1950 DA may be composed of a nickel-iron or enstatite chondritic composition \citep{Busch2007}, and $R$ = 649 m, the mean radius of 1950 DA, we obtain the lowest cohesion to retain the boulder on the surface as 
\begin{eqnarray}
c_{min} &\sim& \frac{2}{3} \rho_G r R \omega^2 - \frac{8 \pi}{9} \rho_G \rho_B G r R, \\
&=& 0.015 \: \text{[Pa/m]} \: r. \nonumber
\end{eqnarray}
where $r$ is the radius of the boulder. Thus, even if the size of the boulder is on the order of a few hundred meters, the lowest cohesion to retain the boulder is on the order of a few pascals. This implies that since the necessary cohesion that prevents 1950 DA from failing structurally is $\sim 80$ Pa, if this is representative of the cohesion within the regolith, any embedded boulders will be stable \citep{Sanchez2014}. This simplified analysis also supports the primary result in this study that the internal core is more sensitive to structural failure than the surface region. 

Finally, we emphasize our contributions. \cite{Rozitis2014} used Holsapple's limit analysis technique \citep{Holsapple2007} to derive the minimum cohesion of 1950 DA as $44 - 74$ Pa. However, since this technique was based on the upper bound theorem \citep{Chen1988}, the cohesion derived by this technique indicates the lowest limit for the necessary condition only. Our finite element model is capable of giving a more precise failure condition and possible failure modes of an asteroid. We note that for this asteroid only a lower bound on strength can be given, as it is not seen to be failing currently. This is different from the study by \cite{Hirabayashi2014_R3} who obtained both lower and upper limits for the cohesion of P/2013 R3 from its disruption event. In addition, we clearly show that the failure mode for this body is for the interior to fail. This leads to a new prediction for the central core of the body to be underdense. Future analysis will explore this in more detail. 


\acknowledgments
MH deeply appreciates Eleanor Matheson for her dedicated review of grammar in the current paper. MH wishes to thank Keith A. Holsapple at University of Washington and Carlos A. Felippa at University of Colorado for their useful advice about development of a finite element model for plastic deformation. MH also acknowledges Ben Rozitis at University of Tennessee for constructive discussions about their technique used in \cite{Rozitis2014}. Finally, MH appreciates Paul S\'anchez for useful discussion about the interpretation of our study. This research was supported by  grant NNA14AB03A from NASA's SSERVI program. 





\clearpage



\clearpage

\begin{table}
\begin{center}
\caption{Physical properties of 1950 DA (the retrograde model)}
\label{Table:prop}
\begin{tabular}{l l c}
\hline 
Property & Value & Reference \\
\hline
\hline
Volume [km$^3$] & 1.145 & \cite{Busch2007} \\
Spin period [hr] & 2.1216 & \cite{Busch2007} \\
Bulk density [g/cm$^3$] & $1.0 - 1.7$ & \cite{Rozitis2014}  \\
Friction Angle & 35 & \cite{Lambe1969} \\ 
Young's modulus [Pa] &  $1.0 \times 10^7$ & - \\
Poisson's ratio & 0.25 & - \\
\hline
\end{tabular}
\end{center}
\end{table}


\end{document}